\documentclass[a4paper,fleqn,usenatbib,useAMS]{mnras}
\usepackage{graphicx}	
\usepackage{amsmath}	
\usepackage{amssymb}    
\usepackage{amssymb}	
\usepackage{multicol}	
\usepackage{bm}		
\usepackage{pdflscape}	
\usepackage{hyperref}

\usepackage{color}

\newcommand*\mean[1]{\bar{#1}}

\newcommand{\mm}{$M_{200\rm{m}}$}
\newcommand{\rockstar}{{\tt ROCKSTAR}}

\newcommand{\hmsun}{h^{-1}\ {\rm M_{\odot}}}
\newcommand{\hMpc}{h^{-1}\ {\rm Mpc}}

\title[]{Halo Exclusion Criteria Impacts Halo Statistics}
\author[Garcia and Rozo]{Rafael Garc\'ia$^1$\thanks{E-mail: rgarciamar@email.arizona.edu}, Eduardo Rozo$^{1}$
\\
$^{1}$Department of Physics, University of Arizona, Tucson, AZ 85721 
}  
\begin{document}

\maketitle 

\label{firstpage}

\begin{abstract}
Every halo finding algorithm must make a critical yet relatively arbitrary choice: it must decide which structures are parent halos, and which structures are sub-halos of larger halos.  We refer to this choice as {\it percolation}. We demonstrate that the choice of percolation impacts the statistical properties of the resulting halo catalog. Specifically, we modify the halo-finding algorithm \rockstar\ to construct four different halo catalogs from the same simulation data, each with identical mass definitions, but different choice of percolation.  The resulting halos exhibit significant differences in both halo abundance and clustering properties. Differences in the halo mass function reach 10\% for halos of mass $10^{13}\ \hmsun$, larger than the few percent precision necessary for current cluster abundance experiments such as the Dark Energy Survey.  Comparable differences are observed in the large-scale clustering bias, while  differences in the halo--matter correlation function reach 40\% on translinear scales. These effects can bias weak-lensing estimates of cluster masses at a level comparable to the statistical precision of current state-of-the-art experiments.
\end{abstract}


\section{Introduction}
\label{intro}

In the halo model the abundance and distribution of galaxies and clusters are linked to the abundance and distribution of dark matter halos \citep{Cooray-Sheth}.  Predicting the properties of halos requires large computer simulations that map the matter distribution of the Universe.  The output of simulations is then analyzed using a halo finder to find gravitationally bound dark matter structures. 

Every halo finding algorithm makes two critical yet relatively arbitrary choices.  The first has received plenty of attention, and is the definition of halo mass.  Halo mass is typically defined as the mass enclosed within some specific spherical aperture, chosen such that the mean density of the halo within that sphere is equal to some factor of either the critical density or the mass density of the Universe. However, other definitions are also commonly used (e.g. friends-of-friends) \citep[see e.g.][]{Knebe2013}. For this reason, one can find calibrations of the halo mass function for multiple different halo mass definitions \citep{Tinker2008, Bhattacharya2011, McClintock2018}.  The second arbitrary choice has received little attention to date, namely, how a halo finding algorithm decides which structures are parent halos, and which are sub-halos that ``belong'' to a larger halo.  We refer to the criteria for categorizing structures as parent halos vs. sub-halos as {\it percolation} or {\it exclusion criteria}.  There is currently no standard percolation scheme, with different halo finders applying different halo exclusion criteria when constructing halo catalogs.

In this paper we show that the choice of percolation impacts the statistical properties of the resulting halo population at a non-negligible level.  To do so, we modify \rockstar\ \citep{Behroozi2013}, a state-of-the-art halo finding algorithm, to generate halo catalogs with identical mass definitions, but different halo exclusion criteria.  For each such halo catalog we measure the halo mass function, correlation function, and projected density profiles.  By comparing these properties of the resulting halo catalogs to the properties of the fiducial \rockstar\ catalog we quantify the level of systematic uncertainty in current theoretical predictions associated with the choice of percolation algorithm implemented in the construction of the catalog.


\section{Simulation Data}
\label{sec:simdata}

We use a cosmological N-body simulation run using the publicly available code \verb+GADGET2+ \citep{Springel2005}. This simulation is similar to the simulations used for the Aemulus project \citep{deroseetal2018}. Specifically, the simulation has a box size of $1050\ \hMpc$ with $1400^3$ particles and was run using periodic boundary conditions with a force softening scale of $20\ h^{-1} {\rm kpc}$. The cosmology of the simulation we use is $h=0.6704$, $\Omega_m=0.318$, $\Omega_{\Lambda}=0.682$, $\Omega_b=0.049$, $\sigma_8=0.835$, $n_s=0.962$. The particle mass is $3.7275 \times 10^{10} \ \hmsun$.  Since the goal of this study is to highlight the under-appreciated impact of halo exclusion on halo statistics, a single simulation suffices for our purposes.

We compare the statistics of halo catalogs generated from the same simulation box, using the same halo mass definition, namely \mm, but different halo exclusion criteria.  The specific statistics we consider are:
\begin{itemize}
    \item the halo mass function,
    \item the large-scale halo clustering bias,
    \item the halo--mass correlation function,
    \item and the halo--halo correlation function.
\end{itemize}
The different halo catalogs are created by modifying the publicly available code \rockstar\ \citep{Behroozi2013}.  

\rockstar\ uses a friends-of-friends algorithm in 6 dimensional phase space to find seed dark matter structures.  It then iteratively assigns particles in the simulation to each seed group, merging seeds into a single halo when the separation between halos is sufficiently small \citep[equation 2 in][]{Behroozi2013}. \rockstar\ then removes all unbound particles from each halo, and computes the spherical mass and radius as defined using a virial overdensity criteria.  Specifically, the mass $M_\Delta$ and spherical radius $R_\Delta$ associated with each halo are selected such that they satisfy the constraint equation 
\begin{equation}
	\frac{4}{3} \pi R_{\Delta}^3 \Delta \mean{\rho}_m = M_{\Delta}.
\end{equation}
Here, $M_\Delta$ is the mass contained within the radius $R_\Delta$, and
$\Delta$ is the overdensity calculated using the spherical collapse model of \citet{BryanNorman1998}.   While the virial overdensity criterion is the default for \rockstar, given the final halo catalog one can readily recompute strict spherical overdensity masses, i.e. masses using all particles, without any unbinding procedure.  In our work, we always define halo mass using strict spherical overdensity masses with a fixed overdensity threshold of $\Delta= 200$ relative to the mean matter density.   Finally, \rockstar\ percolates the seed structure catalog to generate a final halo catalog by determining which seed structures are subhalos of the parent halo centered on a larger substructure.  The classification of a seed structure as a halo or a subhalo is dependent upon the phase-space distance of the seed structure to all larger seeds, and incorporates information from the previous time-step when available.  It is the impact of this percolation step on the statistical properties of the resulting halo catalog which we investigate in this work. 


\section{Percolation Schemes}
\label{percolations}

\begin{figure*}
\centering
\hspace{-12pt} \includegraphics[width=60mm]{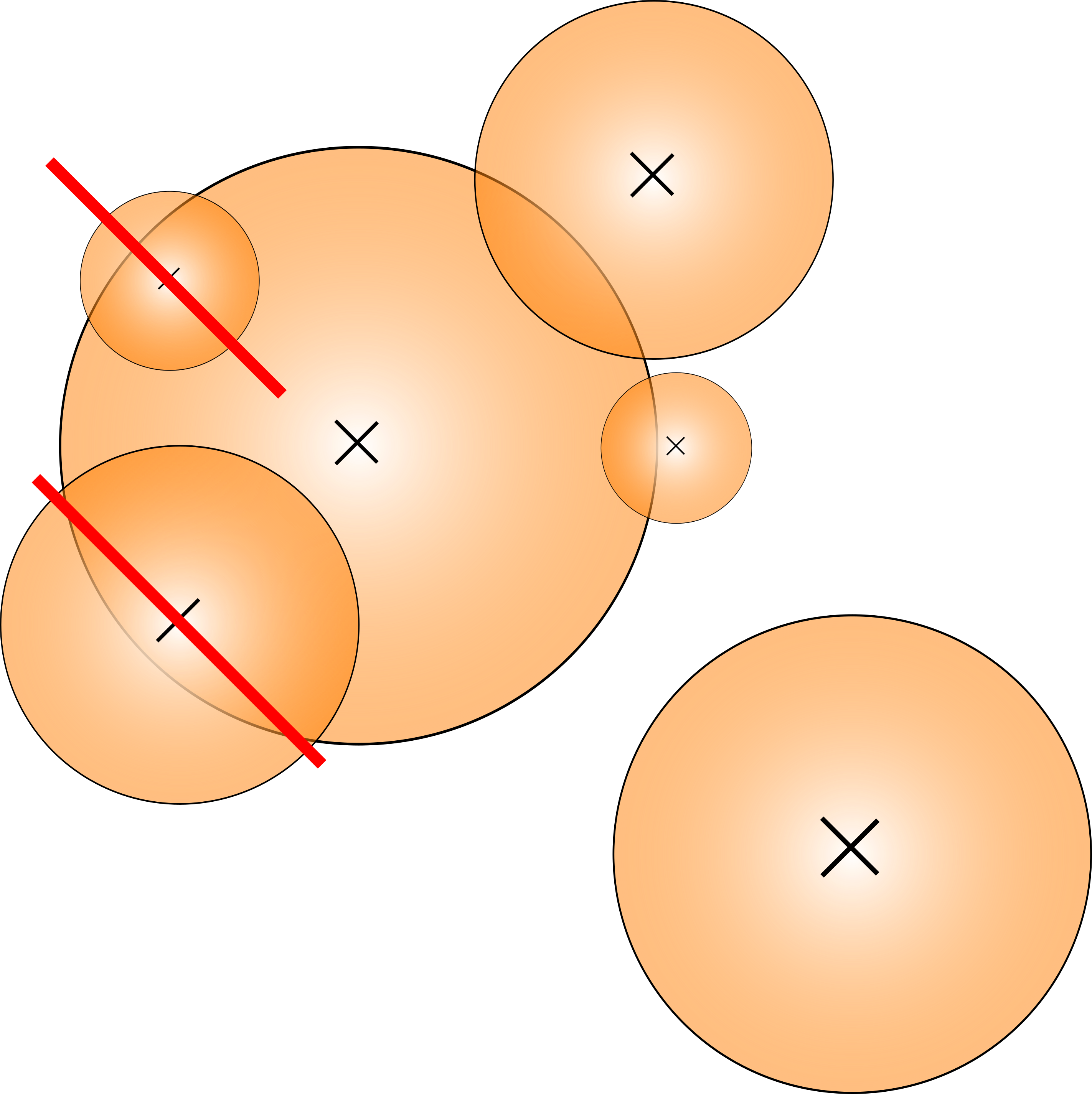}
\hspace{-12pt} \includegraphics[width=60mm]{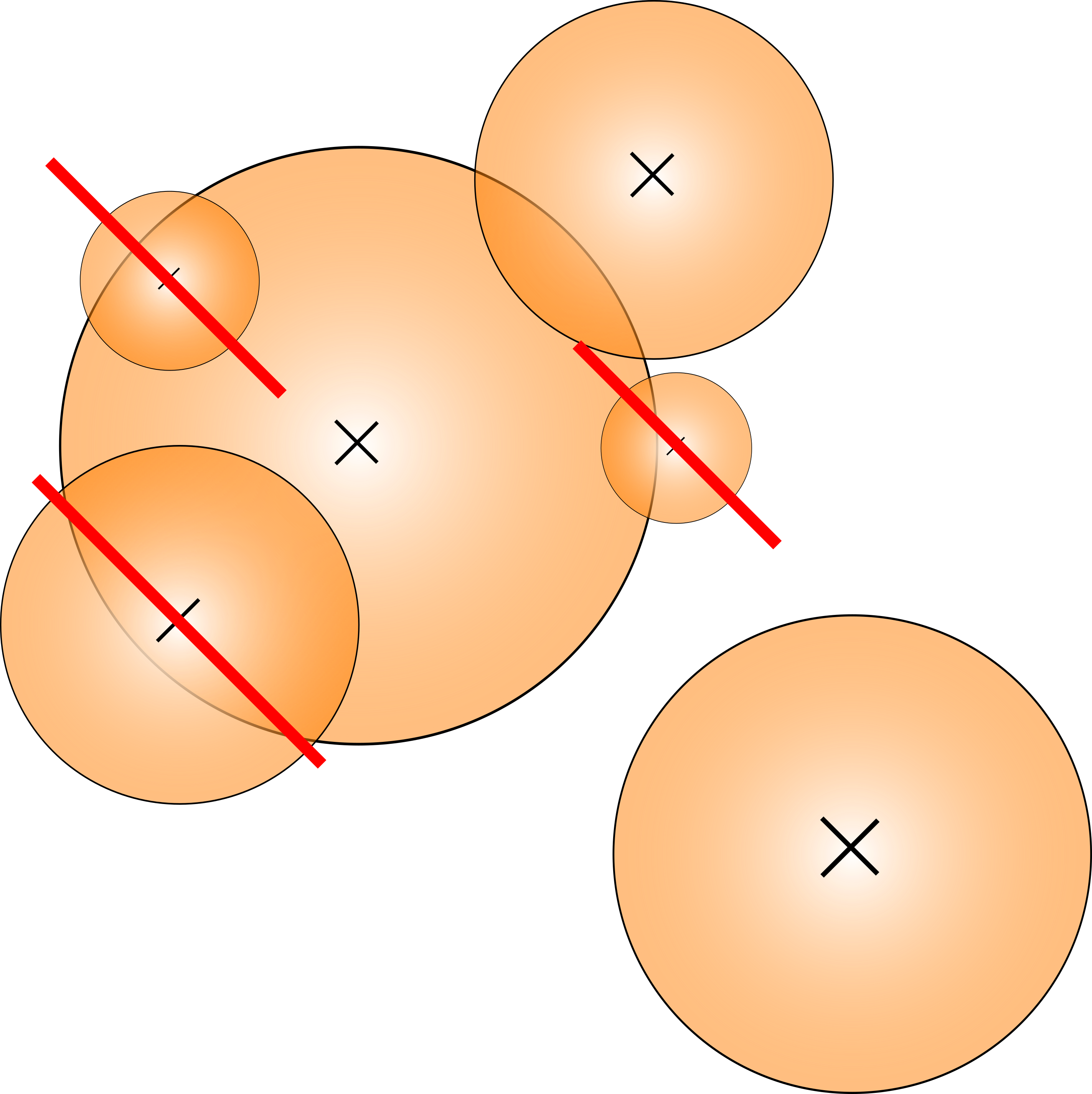}
\hspace{-12pt} \includegraphics[width=60mm]{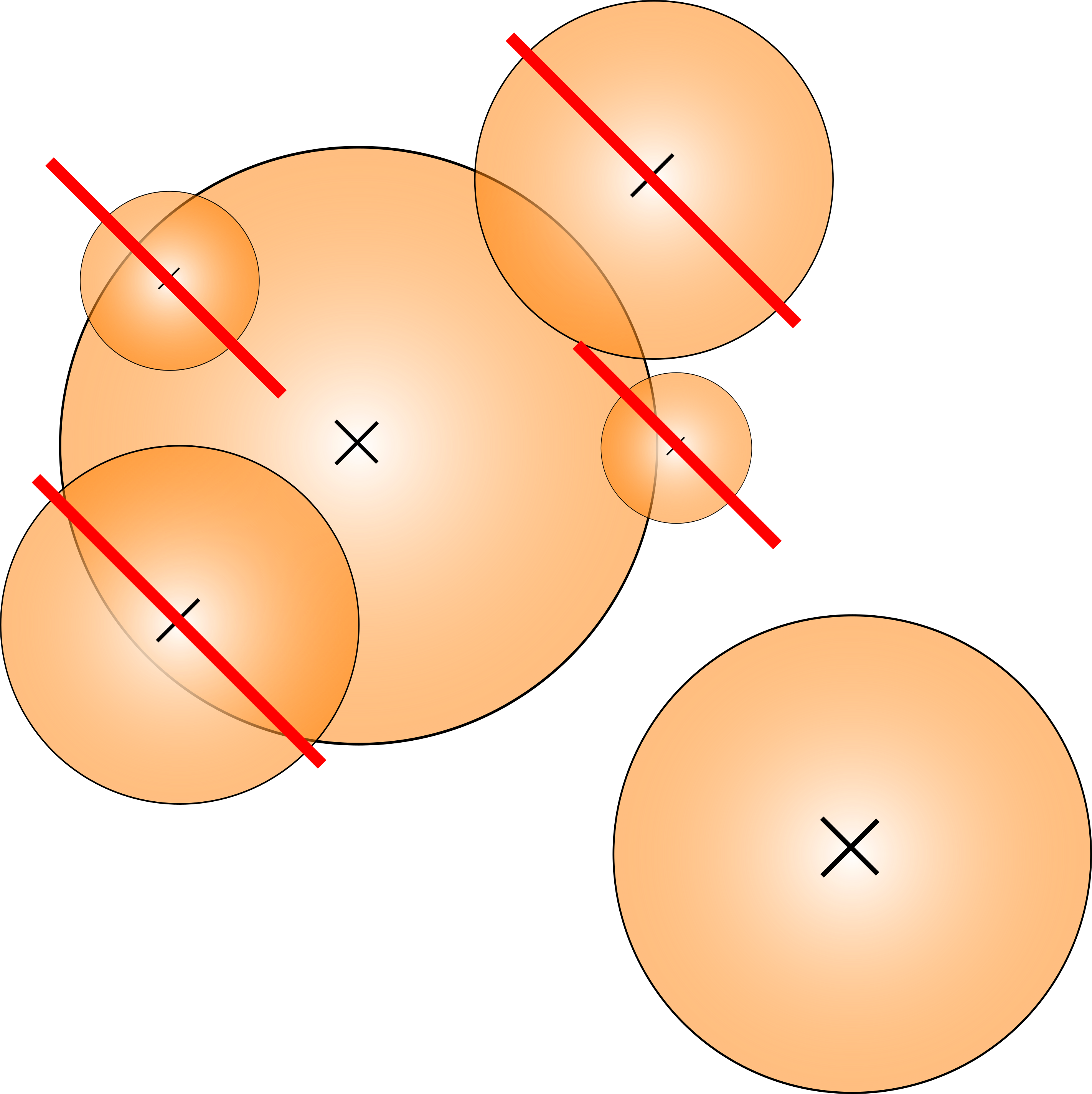} \\
\raggedright
\hspace{40pt} $x_{12} < R_{200} (M_1)$
\hspace{85pt} $x_{12} < R_{200} (M_1 + M_2)$
\hspace{70pt} $x_{12} < R_{200} (M_1) + R_{200} (M_2)$
\caption{Qualitative illustration of different percolation schemes. The red line indicates the removal of a halo seed from the halo catalog. \textbf{Left:} soft sphere scheme, which removes all of the seeds whose center resides within the radius of another halo seed. \rockstar\ percolates in this way, but defining the halo radius as $R_{vir}$, which corresponds to $R_{360m}$ at $z=0$. \textbf{Right:} hard sphere scheme, which removes all of the seeds that overlap with a bigger halo seed. \textbf{Center:} Point-Mass Approximation (PMA) scheme, which takes into account the mass of both halo seeds in question. This scheme allows overlaps between halos but may also remove seeds whose center resides outside of other halos.
}
\label{fig:percschemes}
\end{figure*}

We create alternate halo catalogs starting from the seed structures identified by \rockstar\ by changing the default percolation in the code.  First, we trim the list of seed structures to those above a mass threshold of $M_{200m}\geq 10^{12.5}\ \hmsun$ ($\sim 300$ particles). Seed structures are then classified as halos or sub-halos using a simple spherical exclusion criterion.  Specifically, we rank order all seed structures according to their maximum circular velocity, defined as the maximum of the circular velocity profile
\begin{equation}
V(r) = [GM(<r)/r]^{1/2}.	
\end{equation}

Initially, all seed structures are considered candidate halos.  Starting from the top-ranked (largest) candidate halo, we apply a spherical exclusion criteria to identify substructures of the halo centered on the top-ranked seed structure.  Specifically, given two structures of mass $M_1$ and $M_2$, the two structures are considered to fall within the same parent halo if the separation between the two structures $x_{12}$ satisfies
\begin{equation}
x_{12} \equiv |\bm{r}_1 - \bm{r}_2| \leq d(M_1,M_2) ,
\end{equation} 
where $d$ is the halo exclusion function.  All seed structures identified as substructures of a larger parent halo are removed from the candidate halo list, and the procedure is iterated with the next highest-ranked candidate halo until no more candidate halos remain.  

We consider three different choices for halo exclusion:
\begin{enumerate}
\item \bf Soft-sphere halo exclusion: \rm Two seed structures are considered to be in the same parent halo if their separation is less than the radius of the larger structure, i.e. $d(M_1,M_2) = R_{200}\left( \max(M_1,M_2) \right)$.  This is the halo exclusion criteria used in \citet{Tinker2008}.
\item \bf Point-mass exclusion: \rm Two seed structures are considered to be in the same parent halo if their separation is less than the radius of a structure of mass $M_1+M_2$, i.e. $d(M_1,M_2) = R_{200}(M_1+M_2)$.  This halo exclusion criterion self-consistently enforces strict spherical overdensity mass definitions and exclusion when the halos can be approximated as point particles.
\item \bf Hard-sphere exclusion: \rm Two seed structures are considered to be in the same parent halo if the spherical volumes associated with each structure overlap at all, i.e.  $d(M_1,M_2) = R_{200}(M_1)+R_{200}(M_2)$.
\end{enumerate}        

The three percolation schemes are illustrated in Fig. \ref{fig:percschemes}. There we apply each of our proposed percolations to the same set of halo seeds in an illustrative example in which a large mass halo is surrounded by several smaller halo seeds.  The boundaries we draw are circles of radius $R_{200m} (M)$. The large halo is the top-ranked parent halo. The classification of the rest of the halo seeds as halos or subhalos depends on the percolation scheme: soft-sphere (left), point-mass exclusion (center), or hard-sphere exclusion (right).  The crossed seeds are removed from the final halo catalog and are identified as substructures of the bigger halo in each of the percolation schemes. In terms of the amount of halos removed from the candidate halo list, the soft-sphere scheme (left) is the most conservative scheme, and the hard-sphere scheme (right) is the most aggressive one.


\section{Impact on halo statistics}
\label{sec:results}

We characterize the impact of the percolation scheme used to generate the halo catalog on four different halo statistics: the halo mass function, the halo--matter correlation function, the halo--halo correlation function, and the large-scale clustering bias.  In addition to the three spherical exclusion criteria defined above, we also considered the default \rockstar\ halo catalog.  In all cases, we use strict spherical overdensity masses to define the mass of a halo.  Because the mass definition itself is constant, any differences in the statistics of the four halo catalogs we generated must be the direct result of the different percolation schemes.

\subsection{Halo Mass Function}
\label{sec:hmf}

\begin{figure}
\hspace{-12pt} \includegraphics[width=90mm]{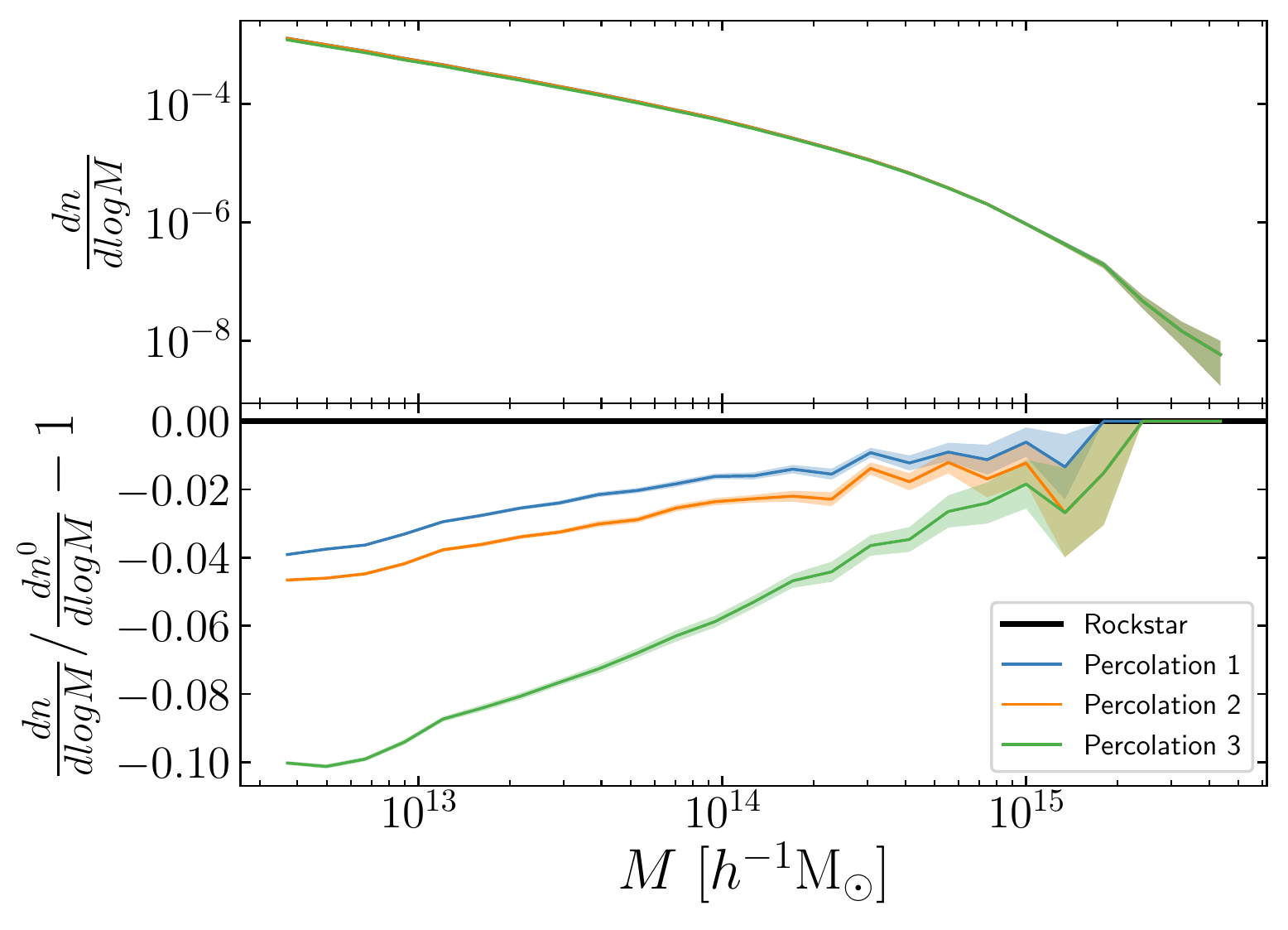}
\caption{The halo mass function for distinct percolation schemes. The choice of percolation has a significant impact on halo abundance. \textbf{Top}: Halo mass functions. \textbf{Bottom:} Fractional difference of the halo mass functions with respect to the fiducial. The shaded regions represent 68\% confidence intervals as determined by jackknifing.}
\label{fig:hmf}
\end{figure}

Figure~\ref{fig:hmf} compares the halo mass functions of the halo catalogs generated using each of the four different percolation algorithms (fiducial, soft-sphere, point-mass, and hard-sphere). The lower panel show the fractional difference in the halo mass function relative to the fiducial percolation.  The impact of percolation is clearly negligible at the high-mass end, but can become significant at low halo masses.  This makes sense. Since high mass halos dominate their environment, the impact of percolation on these halos is negligible: these halos are never assigned as sub-halos of more massive systems. By contrast, the more aggressive percolation schemes remove small halos from the immediate vicinity of large halos, thereby suppressing the resulting halo abundance at low masses.  The largest difference is that between the fiducial \rockstar\ percolation algorithm and the hard-sphere exclusion, reaching $\approx 10\%$ (5\%) differences for mass $M\sim 10^{13}\ \hmsun$ ($M\sim 10^{14}\ \hmsun$) halos.

The differences illustrated in Figure~\ref{fig:hmf} are larger than the $\approx 1\%$ precision necessary for stage III dark energy experiments such as the Dark Energy Survey, and significantly larger than the precision reached by current halo--mass function emulators \citep[e.g.][]{McClintock2018}.  Evidently, while we can make very precise predictions for the halo mass function given a halo-finding algorithm, it is clear that the choice of percolation introduces a significant amount of systematic uncertainty in our predictions.  Moreover, this level of systematic uncertainty is irreducible so long as halos in simulations are percolated in a way that is different from the way clusters are percolated in real data sets.  Our results demonstrate that implementing identical percolation algorithms across both simulated halos and real clusters is necessary for stage III and IV dark energy experiments.


\subsection{Halo--Mass Correlation Function}

\begin{figure*}
\hspace{-12pt} \includegraphics[width=180mm]{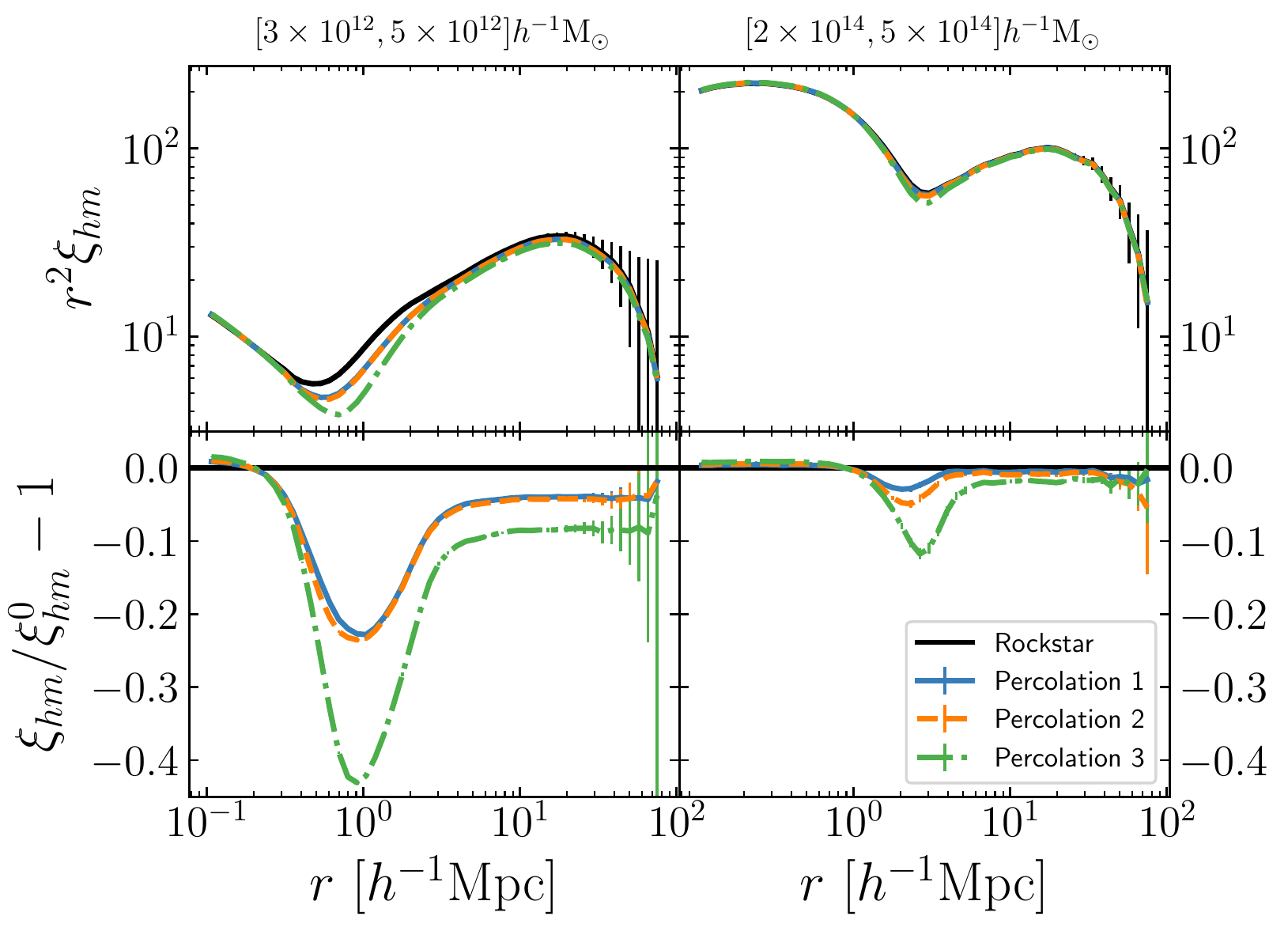}
\caption{Halo-mass correlation function for different percolation schemes and mass bins. \textbf{Top:} Halo-mass correlation function. \textbf{Bottom:} Fractional differences with respect to the fiducial percolation. \textbf{Left:} Low mass halos. \textbf{Right:} High mass halos. Error bars are jackknife.}
\label{fig:xihm}
\end{figure*}


Figure~\ref{fig:xihm} shows the halo--mass correlation function measured for each of our halo catalogs (top row), and the relative difference in the halo--mass correlation functions relative to that measured in our fiducial catalog (bottom row).  The left and right columns correspond to halos with masses in the range $[3\times 10^{12}, 5\times 10^{12}] \hmsun$ and $[2\times 10^{14}, 5\times 10^{14}] \hmsun$ respectively. We plot data for these same mass bins through the rest of this paper. We find that aggressive halo exclusion criteria lead to suppression of the halo-mass correlation function $\xi_{hm}$.  There is an obvious large feature on translinear scales ($\sim 1\ \hMpc$ to a few $\hMpc$), along with a constant change in the clustering amplitude at large scales.  The large (up to 40\% difference) feature at translinear scales makes sense: the more aggressive exclusion criteria remove low-mass halos in the vicinity of high-mass halos.  Consequently, the amount of mass in the immediate neighborhood of the remaining low mass halos is suppressed, leading to a large decrease in the halo--mass correlation function.  The typical length scale associated with this effect is the exclusion radius of the largest dark matter halos.

The fact that there is an overall offset in the clustering amplitude of halos at large scales may seem surprising at first sight.  However, this too is easily explained. A stronger halo exclusion region removes more small halos from the vicinity of large halos.  Since large halos live in high density regions, the surviving low-mass halos must necessarily be less clustered.  As for the halo mass function, these trends are more pronounced for low-mass halos than they are for high-mass halos, and for the same reason: high mass halos dominate their environment, and are therefore rarely removed through percolation.

We characterize the large-scale clustering amplitude in terms of the halo bias, defined as the ratio of the halo--matter cross correlation function and the matter auto-correlation function
\begin{equation}
	b(r|M) = \frac{\xi_{hm} (r|M)} {\xi_{mm} (r)}
\end{equation}
where $M$ is the mass of the halo.  On large scales ($r \geq 10\ \hMpc$), the halo bias is approximately constant, as expected.  We fit a constant bias model to the data in the radial range $10 < r < 80\ \hMpc$ to arrive at our final value for the large-scale bias $b(M)$ in each of our four halo catalogs.  Interestingly, we find that the ratio $\xi_{hm}/\xi_{\rm lin}$, where $\xi_{\rm lin}$ is the linear correlation function, is {\it not} constant over the same scales.  That is, the linear-bias approximation is valid to significantly lower scales provided the reference clustering function is the matter correlation function rather than the linear correlation function.

\begin{figure}
\hspace{-12pt} \includegraphics[width=90mm]{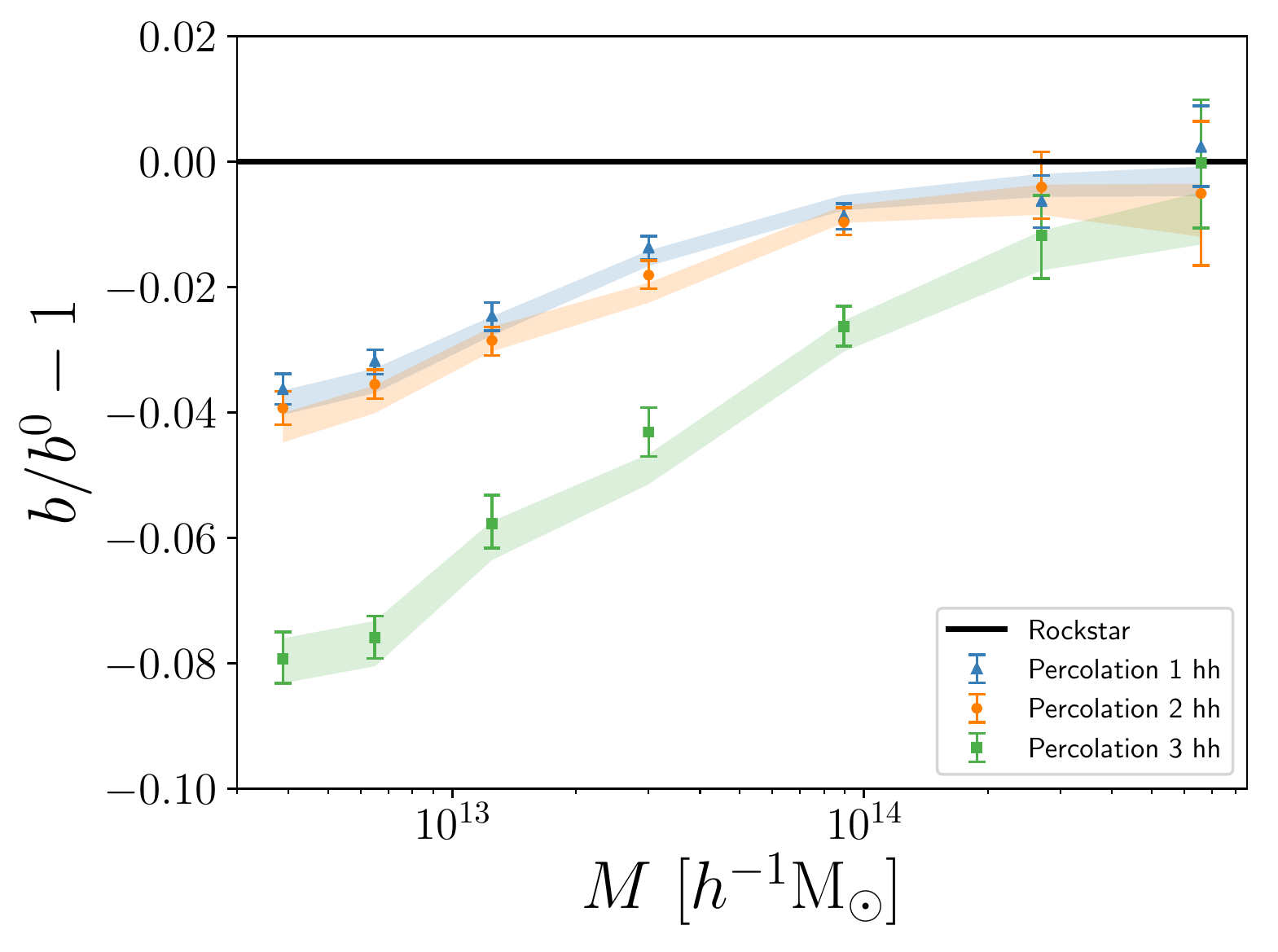}
\caption{Fractional difference between the large scale bias measured by each percolation scheme and the fiducial scheme. Points with error bars represent the bias measured using the halo--halo correlation functions. Colored regions show the bias measured using the halo--mass correlation functions. It is reassuring that the two bias measurements are in excellent agreement with each other.}
\label{fig:biashmdif}
\end{figure}

The left panel in Figure~\ref{fig:biashmdif} shows the fractional difference of the large-scale bias between our different halo catalogs and the fiducial \rockstar\ catalog.  The data points show the relative bias as measured using the halo--halo correlation function (see next section for details), whereas the colored bands represent our measurements using the halo--mass correlation function.  The width of the band is set by the error in our measurement.  As expected, the bias of the high mass halos is largely insensitive to percolation effects, whereas the bias of low-mass ($M\sim 10^{13}\ \hmsun$) halos changes by as much as $\approx 8\%$. 

\begin{figure}
\hspace{-12pt} \includegraphics[width=90mm]{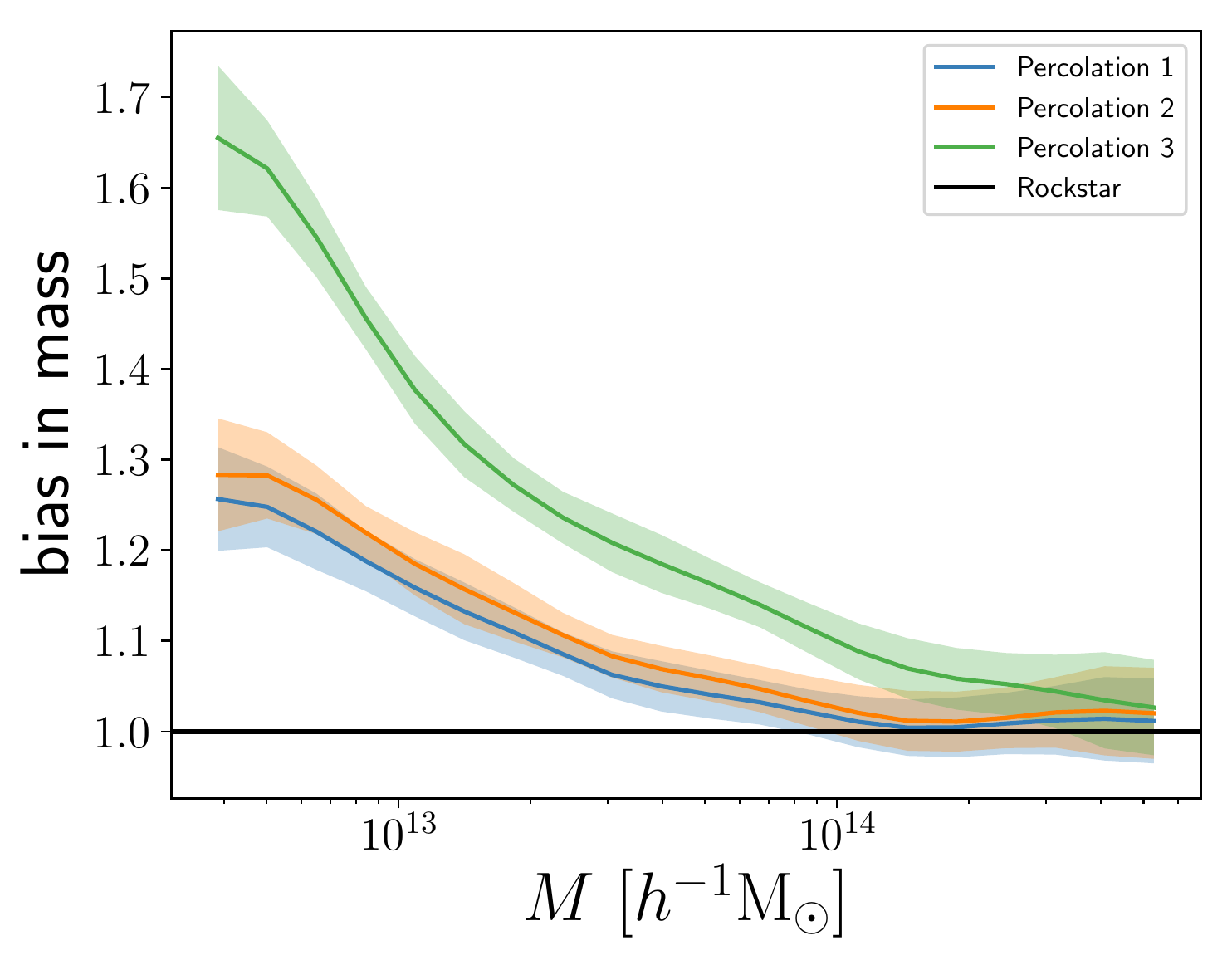}
\caption{Bias in the inferred halo masses of our four catalogs by using large scale clustering amplitude measurements to infer halo masses.  In each case, the clustering amplitude is set by the clustering of the fiducial \rockstar\ halos.  The observed amplitude is then mapped to a new halo mass using each of the alternative halo catalogs in turn.  It is clear that the choice of percolation algorithm can severely impact the inferred halo mass.}
\label{fig:biasmatching}
\end{figure}

The large scale clustering amplitude of cosmological objects is often used as as a way to estimate the mass of the halos hosting those objects \citep[e.g.][]{Robertson2010,Mountrichas2016}.  The dependence of the clustering amplitude on the choice of percolation algorithm demonstrates that these type of estimates can be subject to large systematic uncertainties. To illustrate this, we consider a class of cosmological objects hosted in halos of mass $M$ as defined using the \rockstar\ percolation algorithm.  We calculate the clustering amplitude of these halos, and then use the $b(M)$ relation for the halos in each of our four halo catalogs to infer the corresponding halo mass.  Figure \ref{fig:biasmatching} shows the bias in the inferred halo masses for each of our four halo catalogs.  We see that the choice of percolation algorithm can bias the inferred halo masses by as much as 40\% for halos of mass $M\approx 10^{13}\ \hmsun$.

The translinear regime of the halo--correlation function has long been difficult to model, requiring ad-hoc parameterizations that are then calibrated in numerical simulations \citep[e.g.][]{Surhud2013}.  Such an approach is likely sufficient within the context of modeling galaxy--galaxy clustering for the cosmological purposes, though we caution that verifying robustness of the modeling to simulations populated based on a different halo definition would be worthwhile.  At the very least, inferences about how galaxy populate halos will necessarily be impacted by the differences highlighted above.  The sensitivity to halo definition will be even more problematic for cosmological studies that rely on the halo statistics directly, e.g. cluster abundance studies.  We emphasize again that the work here is not meant to \it calibrate \rm this effect, but rather to demonstrate its existence.  Calibrations for observational studies must be specifically tailored to the observational methodologies employed.

Finally, our results bear some impact on the location of the splashback radius as found both in simulations and data. Specifically, the sharp steepening in the halo--mass correlation that occurs at the translinear regime has been identified with the splashback radius, the distance to the apocenter of dark matter substructures falling into a dark matter halo after their first pericenter pass \citep{adhikari2014}.  This splashback radius has been proposed as a physical definition for the halo boundary \citep{diemerkravtsov14, More2015}. As demonstrated in this work, the steepening feature of the stacked halo profiles for halos of a given mass can be moved by a change in the choice of halo percolation.  This is not in itself problematic: in changing the population of halos being stacked, the distribution of mass accretion rates of the resulting halos will likely change, which in turn will move the average splashback radius \citep{More2015}. It does demonstrate, however, that calibration of the splashback radius via halo stacking is prone to systematics arising from the choice of halo percolation.  This is particularly true for measurements of the splashback radius based on observationally selected cluster samples \citep[e.g.][]{moreetal2016,buschwhite17,umetsuetal2017,baxteretal2017,shinetal2018,zuercheretal2018,changetal2018,contigianietal2019}. Splashback measurements based on the analysis of particle orbits are, of course, free of such systematics \citep{diemer2017,diemeretal2017}.

\subsection{Halo--Halo Correlation Function}


We computed the halo auto-correlation functions for the same mass bins for which we calculated the halo--mass correlation function.  As before, the auto correlation functions exhibit an overall decrease in clustering amplitude for more aggressive halo exclusion criteria.  As in the case for the halo--mass correlation function, we see the appearance of features in the translinear regime, though these features are less apparent than for the case of the halo--mass correlation function: even the translinear feature in the autocorrelation of our lowest mass bins has an amplitude of only $\sim 10\%$. Sample halo--halo correlation function plots are shown in Appendix~\ref{app:plots}.

In a way analogous to the halo--mass correlation function, we can define the large scale halo bias via
\begin{equation}
	b^2(r|M) = \frac{\xi_{hh} (r|M)} {\xi_{mm} (r)}.
\end{equation}
We fit a constant halo bias model over the same radial range as employed in our analysis of the halo--mass correlation function ($r\in[10,80]\ \hMpc$).  The data points in the left panel of Figure~\ref{fig:biashmdif} show the change in the clustering bias relative to our fiducial measurement for each of our four halo catalogs.  We see that the change in the clustering bias amplitude is consistent across the halo--mass and halo--halo correlation function measurements.

\section{Summary and Conclusions}

We have shown that halo exclusion criteria impact halo statistics. Specifically, we modified the percolation of the \rockstar\ halo finding algorithm to generate four different halo catalogs with four different halo exclusion criteria, but identical mass definitions. We then measured the halo mass functions, halo-matter and halo-halo correlation functions, and large-scale clustering bias of the resulting catalogs.  We compared these statistics to the halo statistics of the fiducial halo catalog to quantify the level of uncertainty that the thus-far arbitrary choice of halo percolation scheme introduces in halo clustering statistics. 

We find:
\begin{itemize}
\item The choice of halo exclusion criteria introduces a significant amount of systematic uncertainty on the halo mass function. The largest difference observed in this work was $\approx 10\%$ at a halo mass scale $M\sim 10^{13} \hmsun$.  This value corresponds to the difference between the fiducial and hard-sphere percolation schemes.  Consequently, it is of critical importance for future work on cluster abundances to implement identical exclusion criteria in both theory and simulations, particularly as the low mass threshold for cluster detection gets progressively lower.  Notably, {\it none} of the current choices of halo percolation can be implemented observationally. 

\item The choice of percolation impacts the halo-matter correlations in two ways. At intermediate ($\approx 1\ \hMpc$) scales, large ($\approx 40\%$) relative differences in the halo-mass correlation function of the different halo catalogs arise.  In addition, at large scales we see an offset in the large scale clustering bias.  We demonstrate that halo mass estimates based on the clustering amplitude of a set of cosmological objects can be biased by as much as 40\% due to the choice of percolation used when calibrating the bias--mass relation for halos.

\end{itemize}

The differences in the predicted halo--mass correlation function of halos will necessarily propagate into the predicted weak lensing profiles of the resulting halo population, leading to further sources of systematic uncertainty impacting cluster abundance studies, a systematic which we intend to quantify in future work.  

It is worth noting that while these differences are similar in spirit to differences associated with halo mass definition, a ``right'' answer would naively appear to be more elusive.  Within the context of halo mass definitions, the splashback radius \citep[e.g.][]{More2015} is now a leading contender for the ``right'' radius at which to define the halo edge, naturally leading one to define halo mass as the mass contained within the splashback radius of a halo.  By contrast, no similar leading candidate exists within the context of percolation schemes.  We note, however, that the point--mass exclusion criterion adopted in this work is clearly the one closest in spirit to that of a strict spherical-overdensity mass definition.  We will investigate in future work whether one can define objective quantitative criteria that might lead one to select one exclusion criteria over another within the context of specific science goals.

{\it Acknowledgements:} ER is supported by DOE grant DE-SC0015975 and the Cottrell Scholar Award program. We would also like to thank Risa Wechsler, Joe de Rose, and Matt Becker for making the simulation used in this study available to us, Tom McClintock for technical help throughout the project, and Surhud More, Matt Becker, and Tom McClintock for comments on an early version of this manuscript.

\bibliographystyle{mnras}
\bibliography{database.bib} 

\begin{thebibliography}{}
\makeatletter
\relax
\def\mn@urlcharsother{\let\do\@makeother \do\$\do\&\do\#\do\^\do\_\do\%\do\~}
\def\mn@doi{\begingroup\mn@urlcharsother \@ifnextchar [ {\mn@doi@}
  {\mn@doi@[]}}
\def\mn@doi@[#1]#2{\def\@tempa{#1}\ifx\@tempa\@empty \href
  {http://dx.doi.org/#2} {doi:#2}\else \href {http://dx.doi.org/#2} {#1}\fi
  \endgroup}
\def\mn@eprint#1#2{\mn@eprint@#1:#2::\@nil}
\def\mn@eprint@arXiv#1{\href {http://arxiv.org/abs/#1} {{\tt arXiv:#1}}}
\def\mn@eprint@dblp#1{\href {http://dblp.uni-trier.de/rec/bibtex/#1.xml}
  {dblp:#1}}
\def\mn@eprint@#1:#2:#3:#4\@nil{\def\@tempa {#1}\def\@tempb {#2}\def\@tempc
  {#3}\ifx \@tempc \@empty \let \@tempc \@tempb \let \@tempb \@tempa \fi \ifx
  \@tempb \@empty \def\@tempb {arXiv}\fi \@ifundefined
  {mn@eprint@\@tempb}{\@tempb:\@tempc}{\expandafter \expandafter \csname
  mn@eprint@\@tempb\endcsname \expandafter{\@tempc}}}

\bibitem[\protect\citeauthoryear{{Adhikari}, {Dalal}  \&
  {Chamberlain}}{{Adhikari} et~al.}{2014}]{adhikari2014}
{Adhikari} S.,  {Dalal} N.,   {Chamberlain} R.~T.,  2014, \mn@doi [Journal of
  Cosmology and Astro-Particle Physics] {10.1088/1475-7516/2014/11/019}, \href
  {https://ui.adsabs.harvard.edu/\#abs/2014JCAP...11..019A} {2014, 019}

\bibitem[\protect\citeauthoryear{{Baxter} et~al.,}{{Baxter}
  et~al.}{2017}]{baxteretal2017}
{Baxter} E.,  et~al., 2017, \mn@doi [\apj] {10.3847/1538-4357/aa6ff0}, \href
  {https://ui.adsabs.harvard.edu/\#abs/2017ApJ...841...18B} {841, 18}

\bibitem[\protect\citeauthoryear{{Behroozi}, {Wechsler}  \& {Wu}}{{Behroozi}
  et~al.}{2013}]{Behroozi2013}
{Behroozi} P.~S.,  {Wechsler} R.~H.,   {Wu} H.-Y.,  2013, \mn@doi [\apj]
  {10.1088/0004-637X/762/2/109}, \href
  {http://adsabs.harvard.edu/abs/2013ApJ...762..109B} {762, 109}

\bibitem[\protect\citeauthoryear{{Bhattacharya}, {Heitmann}, {White},
  {Luki{\'c}}, {Wagner}  \& {Habib}}{{Bhattacharya}
  et~al.}{2011}]{Bhattacharya2011}
{Bhattacharya} S.,  {Heitmann} K.,  {White} M.,  {Luki{\'c}} Z.,  {Wagner} C.,
   {Habib} S.,  2011, \mn@doi [\apj] {10.1088/0004-637X/732/2/122}, \href
  {http://adsabs.harvard.edu/abs/2011ApJ...732..122B} {732, 122}

\bibitem[\protect\citeauthoryear{{Bryan} \& {Norman}}{{Bryan} \&
  {Norman}}{1998}]{BryanNorman1998}
{Bryan} G.~L.,  {Norman} M.~L.,  1998, \mn@doi [\apj] {10.1086/305262}, \href
  {http://adsabs.harvard.edu/abs/1998ApJ...495...80B} {495, 80}

\bibitem[\protect\citeauthoryear{{Busch} \& {White}}{{Busch} \&
  {White}}{2017}]{buschwhite17}
{Busch} P.,  {White} S.~D.~M.,  2017, \mn@doi [\mnras] {10.1093/mnras/stx1584},
  \href {http://adsabs.harvard.edu/abs/2017MNRAS.470.4767B} {470, 4767}

\bibitem[\protect\citeauthoryear{{Chang} et~al.,}{{Chang}
  et~al.}{2018}]{changetal2018}
{Chang} C.,  et~al., 2018, \mn@doi [\apj] {10.3847/1538-4357/aad5e7}, \href
  {https://ui.adsabs.harvard.edu/\#abs/2018ApJ...864...83C} {864, 83}

\bibitem[\protect\citeauthoryear{{Contigiani}, {Hoekstra}  \&
  {Bah{\'e}}}{{Contigiani} et~al.}{2019}]{contigianietal2019}
{Contigiani} O.,  {Hoekstra} H.,   {Bah{\'e}} Y.~M.,  2019, \mn@doi [\mnras]
  {10.1093/mnras/stz404}, \href
  {https://ui.adsabs.harvard.edu/\#abs/2019MNRAS.tmp..411C} {p.~411}

\bibitem[\protect\citeauthoryear{{Cooray} \& {Sheth}}{{Cooray} \&
  {Sheth}}{2002}]{Cooray-Sheth}
{Cooray} A.,  {Sheth} R.,  2002, \mn@doi [\physrep]
  {10.1016/S0370-1573(02)00276-4}, \href
  {http://adsabs.harvard.edu/abs/2002PhR...372....1C} {372, 1}

\bibitem[\protect\citeauthoryear{{DeRose} et~al.,}{{DeRose}
  et~al.}{2018}]{deroseetal2018}
{DeRose} J.,  et~al., 2018, arXiv e-prints, \href
  {https://ui.adsabs.harvard.edu/\#abs/2018arXiv180405865D} {p.
  arXiv:1804.05865}

\bibitem[\protect\citeauthoryear{{Diemer}}{{Diemer}}{2017}]{diemer2017}
{Diemer} B.,  2017, \mn@doi [\apjs] {10.3847/1538-4365/aa799c}, \href
  {http://adsabs.harvard.edu/abs/2017ApJS..231....5D} {231, 5}

\bibitem[\protect\citeauthoryear{{Diemer} \& {Kravtsov}}{{Diemer} \&
  {Kravtsov}}{2014}]{diemerkravtsov14}
{Diemer} B.,  {Kravtsov} A.~V.,  2014, \mn@doi [\apj]
  {10.1088/0004-637X/789/1/1}, \href
  {http://adsabs.harvard.edu/abs/2014ApJ...789....1D} {789, 1}

\bibitem[\protect\citeauthoryear{{Diemer}, {Mansfield}, {Kravtsov}  \&
  {More}}{{Diemer} et~al.}{2017}]{diemeretal2017}
{Diemer} B.,  {Mansfield} P.,  {Kravtsov} A.~V.,   {More} S.,  2017, \mn@doi
  [\apj] {10.3847/1538-4357/aa79ab}, \href
  {http://adsabs.harvard.edu/abs/2017ApJ...843..140D} {843, 140}

\bibitem[\protect\citeauthoryear{{Knebe} et~al.,}{{Knebe}
  et~al.}{2013}]{Knebe2013}
{Knebe} A.,  et~al., 2013, \mn@doi [\mnras] {10.1093/mnras/stt1403}, \href
  {http://adsabs.harvard.edu/abs/2013MNRAS.435.1618K} {435, 1618}

\bibitem[\protect\citeauthoryear{{McClintock} et~al.,}{{McClintock}
  et~al.}{2019}]{McClintock2018}
{McClintock} T.,  et~al., 2019, \mn@doi [\apj] {10.3847/1538-4357/aaf568},
  \href {https://ui.adsabs.harvard.edu/\#abs/2019ApJ...872...53M} {872, 53}

\bibitem[\protect\citeauthoryear{{More}, {Diemer}  \& {Kravtsov}}{{More}
  et~al.}{2015}]{More2015}
{More} S.,  {Diemer} B.,   {Kravtsov} A.~V.,  2015, \mn@doi [\apj]
  {10.1088/0004-637X/810/1/36}, \href
  {https://ui.adsabs.harvard.edu/\#abs/2015ApJ...810...36M} {810, 36}

\bibitem[\protect\citeauthoryear{{More} et~al.,}{{More}
  et~al.}{2016}]{moreetal2016}
{More} S.,  et~al., 2016, \mn@doi [\apj] {10.3847/0004-637X/825/1/39}, \href
  {https://ui.adsabs.harvard.edu/\#abs/2016ApJ...825...39M} {825, 39}

\bibitem[\protect\citeauthoryear{{Mountrichas}, {Georgakakis}, {Menzel},
  {Fanidakis}, {Merloni}, {Liu}, {Salvato}  \& {Nandra}}{{Mountrichas}
  et~al.}{2016}]{Mountrichas2016}
{Mountrichas} G.,  {Georgakakis} A.,  {Menzel} M.~L.,  {Fanidakis} N.,
  {Merloni} A.,  {Liu} Z.,  {Salvato} M.,   {Nandra} K.,  2016, \mn@doi
  [\mnras] {10.1093/mnras/stw281}, \href
  {https://ui.adsabs.harvard.edu/\#abs/2016MNRAS.457.4195M} {457, 4195}

\bibitem[\protect\citeauthoryear{{Robertson}}{{Robertson}}{2010}]{Robertson2010}
{Robertson} B.~E.,  2010, \mn@doi [\apj] {10.1088/2041-8205/716/2/L229}, \href
  {https://ui.adsabs.harvard.edu/\#abs/2010ApJ...716L.229R} {716, L229}

\bibitem[\protect\citeauthoryear{{Shin} et~al.,}{{Shin}
  et~al.}{2018}]{shinetal2018}
{Shin} T.,  et~al., 2018, arXiv e-prints, \href
  {https://ui.adsabs.harvard.edu/\#abs/2018arXiv181106081S} {p.
  arXiv:1811.06081}

\bibitem[\protect\citeauthoryear{{Springel}}{{Springel}}{2005}]{Springel2005}
{Springel} V.,  2005, \mn@doi [\mnras] {10.1111/j.1365-2966.2005.09655.x},
  \href {http://adsabs.harvard.edu/abs/2005MNRAS.364.1105S} {364, 1105}

\bibitem[\protect\citeauthoryear{{Tinker}, {Kravtsov}, {Klypin}, {Abazajian},
  {Warren}, {Yepes}, {Gottl{\"o}ber}  \& {Holz}}{{Tinker}
  et~al.}{2008}]{Tinker2008}
{Tinker} J.,  {Kravtsov} A.~V.,  {Klypin} A.,  {Abazajian} K.,  {Warren} M.,
  {Yepes} G.,  {Gottl{\"o}ber} S.,   {Holz} D.~E.,  2008, \mn@doi [\apj]
  {10.1086/591439}, \href {http://adsabs.harvard.edu/abs/2008ApJ...688..709T}
  {688, 709}

\bibitem[\protect\citeauthoryear{{Umetsu} \& {Diemer}}{{Umetsu} \&
  {Diemer}}{2017}]{umetsuetal2017}
{Umetsu} K.,  {Diemer} B.,  2017, \mn@doi [\apj] {10.3847/1538-4357/aa5c90},
  \href {https://ui.adsabs.harvard.edu/\#abs/2017ApJ...836..231U} {836, 231}

\bibitem[\protect\citeauthoryear{{Zuercher} \& {More}}{{Zuercher} \&
  {More}}{2018}]{zuercheretal2018}
{Zuercher} D.,  {More} S.,  2018, arXiv e-prints, \href
  {https://ui.adsabs.harvard.edu/\#abs/2018arXiv181106511Z} {p.
  arXiv:1811.06511}

\bibitem[\protect\citeauthoryear{{van den Bosch}, {More}, {Cacciato}, {Mo}  \&
  {Yang}}{{van den Bosch} et~al.}{2013}]{Surhud2013}
{van den Bosch} F.~C.,  {More} S.,  {Cacciato} M.,  {Mo} H.,   {Yang} X.,
  2013, \mn@doi [\mnras] {10.1093/mnras/sts006}, \href
  {https://ui.adsabs.harvard.edu/\#abs/2013MNRAS.430..725V} {430, 725}

\makeatother
\end{thebibliography}


\appendix

\section{Sample Halo--Halo Correlation Function Plots}
\label{app:plots}

\begin{figure*}
\hspace{-12pt} \includegraphics[width=180mm]{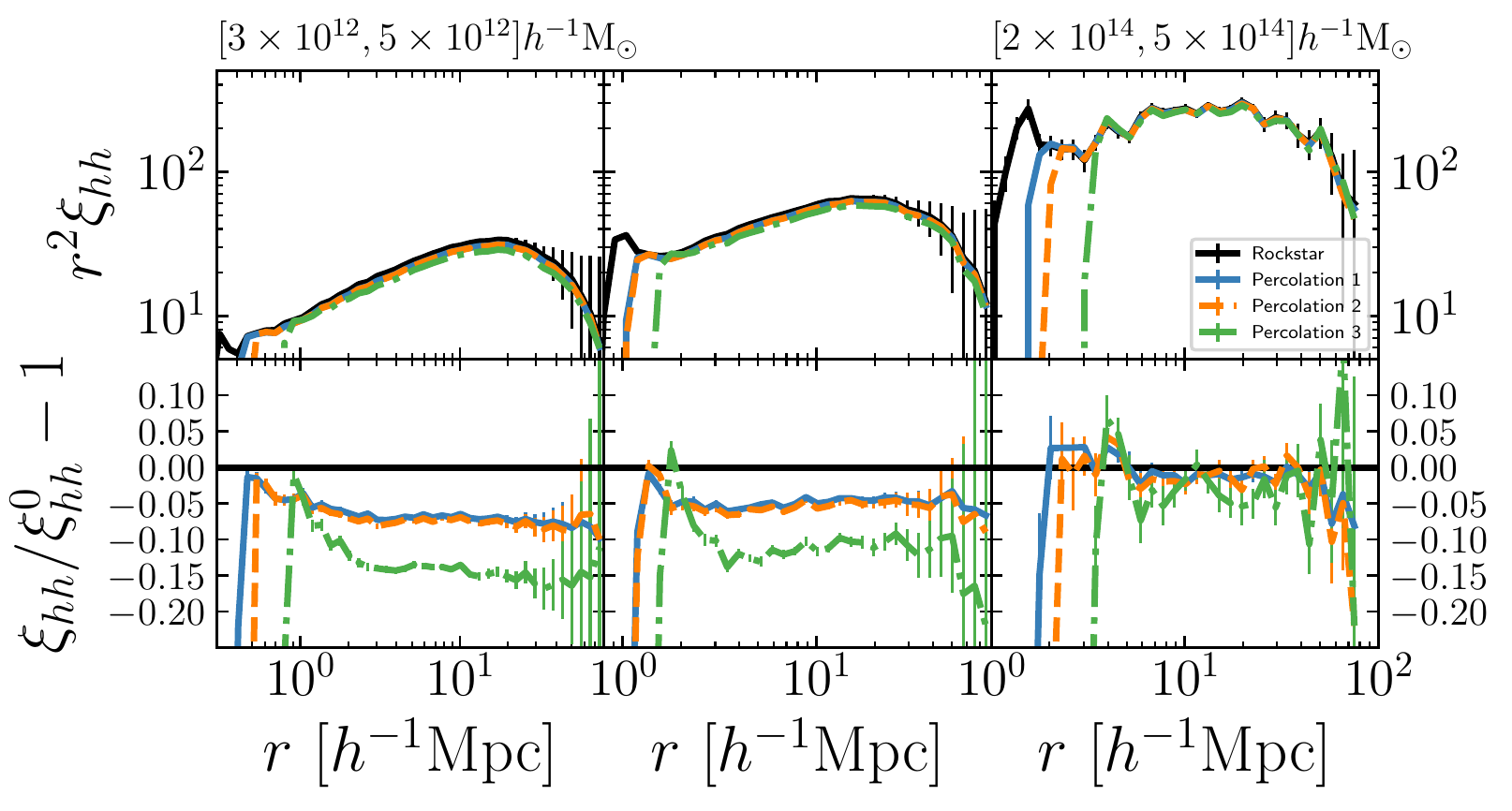}
\caption{The halo--halo auto and cross correlation functions for the two
mass bins used throughout the paper, as labeled.  The top row shows the correlation functions
for each of our four halo catalogs, whereas the bottom row shows the difference relative to the
the correlation function for the fiducial \rockstar\ halo catalog.  The left-most column corresponds
to our low-mass bin, the right-most column to the high-mass bin, and the central column shows
the cross-correlation between the two.}
\label{fig:xihh}
\end{figure*}

Figure~\ref{fig:xihh} shows the halo--halo auto and cross correlation functions for the two
mass bins used throughout the paper, as labeled.  These plots are included here for completeness.

\label{lastpage}

\end{document}